\documentclass[12pt]{iopart}

\usepackage{iopams}
\usepackage{graphicx}
\usepackage[numbers,sort&compress]{natbib}
\begin{document}

\title{Reliable transport through a microfabricated X-junction surface-electrode ion trap}
\author{Kenneth Wright\footnote{Present address: Dept. of Physics, University of Maryland}, Jason M Amini\footnote{Author to whom correspondence should be addressed}, Daniel L Faircloth\footnote{Present address: Ierus Technologies, 9122 Loxford Street, Lithia Springs, GA 30122}, Curtis Volin, S Charles Doret, Harley Hayden, C-S Pai, David W Landgren, Douglas Denison,  Tyler Killian\textmd{$||$},  Richart E Slusher, and Alexa W Harter}
\address{Georgia Tech Research Institute, Atlanta, GA  30332}
\ead{jason.amini@gtri.gatech.edu}
\date{\today}

\begin{abstract}
We report the design, fabrication, and characterization of a microfabricated surface-electrode ion trap that supports controlled transport through the two-dimensional intersection of linear trapping zones arranged in a ninety-degree cross. The trap is fabricated with very-large-scalable-integration (VLSI) techniques which are compatible with scaling to a large quantum information processor. The shape of the radio-frequency (RF) electrodes is optimized with a genetic algorithm to reduce axial pseudopotential barriers and minimize ion heating during transport. Seventy-eight independent DC control electrodes enable fine control of the trapping potentials. We demonstrate reliable ion transport between junction legs and determine the rate of ion loss due to transport.  Doppler-cooled ions survive more than 10$^5$ round-trip transits between junction legs without loss and more than sixty-five consecutive round trips without laser cooling.

\end{abstract}

\maketitle

\newcommand{\shorttitle}{Transport through a microfabricated X-junction}

\tableofcontents

\renewcommand{\leftmark}{\shorttitle}

\section{Introduction}
Trapped atomic ions are one of the most promising systems yet proposed for large-scale quantum information processing (QIP) and quantum simulation~\cite{Ladd2010}. Trapped ions benefit from long coherence times and have been used to perform high-fidelity single- and two-qubit gates~\cite{H.Haeffner2008, R.Blatt2008, Wineland2011, JulioT.Barreiro12011, R.Blatt2012}.  As systems grow to larger numbers of ions, however, ion traps will require new features to facilitate experiments.  Most notably, existing proposals for constructing a large ion-trap quantum computer call for junction elements to manipulate ion positions within the trap.  For example, one proposal~\cite{D.Kielpinski2002} would arrange many ion traps in a two-dimensional array, with junctions shuttling ions between separated computation and storage zones.  A second proposal~\cite{S.Korenblit2012, Lin2009} envisions co-trapping two species of ions in long chains, using one species for QIP and the second for sympathetic cooling.  Since such dual-species chains are cooled most efficiently when ions are ordered in a particular way~\cite{Duan2011}, junctions would be required to establish and preserve the correct sequence of ions.  

Junction ion traps have been previously demonstrated by several groups, beginning with multi-substrate T-~\cite{Hensinger2006} and X-junctions~\cite{Blakestad2009}, and reliable transport with sub-phonon motional heating has been demonstrated in the X-junction~\cite{Blakestad2011}.  However, such multi-substrate constructon is not amenable to scaling to larger systems, making these traps impractical for large-scale QIP.  Fortunately, new generations of microfabricated ion traps -- particularly surface-electrode traps -- provide an attractive alternative~\cite{Hughes2011, J.Chiaverini2005}.  Microfabricated traps may be built using scalable methods, and their small feature sizes permit electrode designs that offer unprecedented control over trapped-ion positions.  Scalable microfabricated junction traps have recently been demonstrated~\cite{Amini2010, Moehring2011}, but transport through these junctions in the absence of Doppler cooling has not been systematically studied, thus it is unknown if the heating in these traps is low enough to support quantum information processing.

Here we report the design, fabrication, and characterization of a surface-electrode X-junction ion trap fabricated with standard VLSI-compatible processes and suitable for use in a large-scale quantum information processor.  In contrast to previous work with microfabricated junction traps, we perform a detailed study of ion loss induced by transporting ions between legs of the junction, both with and without Doppler cooling.  We characterize ion transport through the junction \emph{with cooling} by performing 10$^6$ shuttling operations, determining statistical bounds on the transport fidelity. We find motional heating to be sufficiently low that ions may be shuttled through the junction without Doppler cooling more than sixty-five times without loss.

\section{Design and fabrication}
\begin{figure}
\center 
\includegraphics[scale=1]{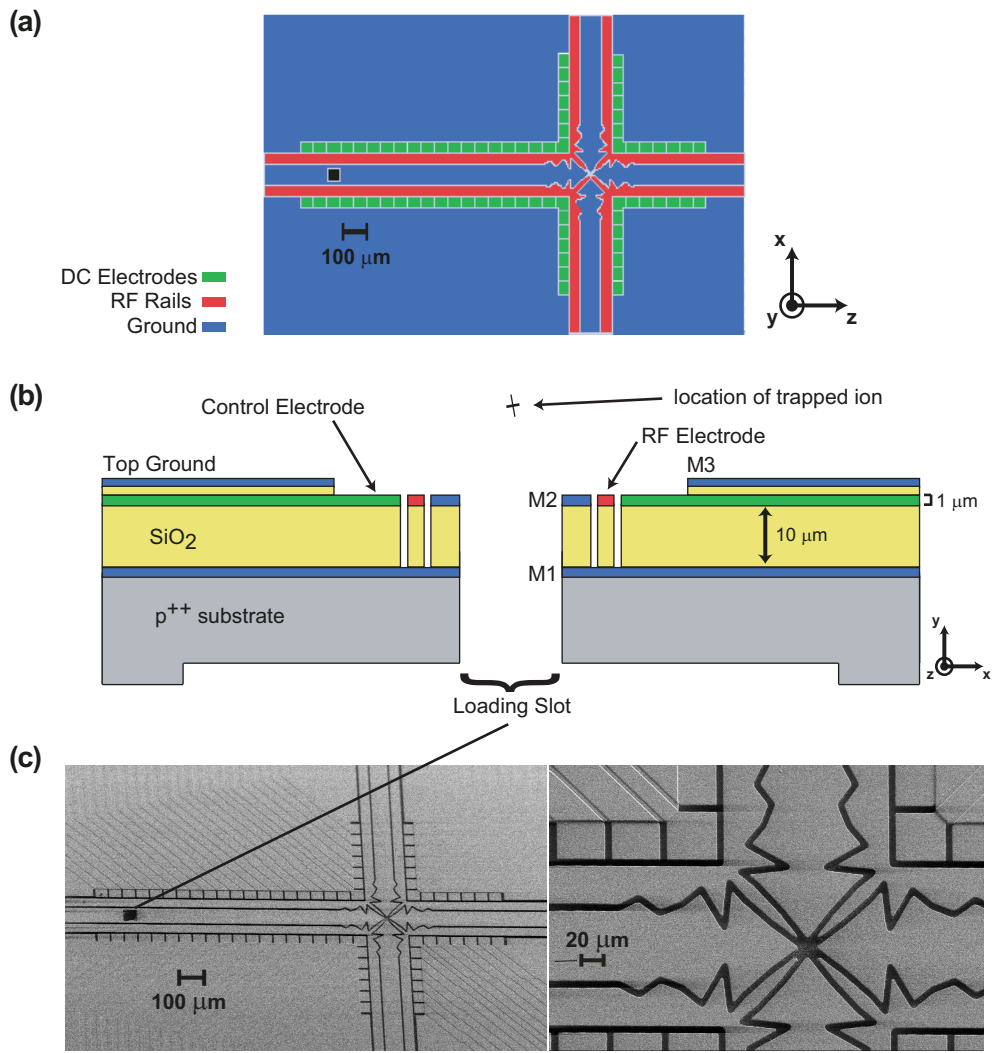} 
\caption{(a) View of the trap from above, showing the control electrodes (green), the RF electrodes (red), and the top-level ground and grounded center electrodes (blue). (b) Cross section of the trap (not to scale). Metal (aluminum) layers are denoted M1 (bottom ground plane), M2 (RF and control electrodes, filter capacitors, and wire bond pads), and M3 (top-level ground plane).  (c) Scanning electron micrographs (SEMs) of the completed trap, including a closeup of the junction center.}
\label{fig:MetalLayers}
\end{figure}

The trap's basic design is that of a symmetric five-wire geometry~\cite{J.Chiaverini2005} with segmented outer control electrodes (figure 1a). Its internal layer structure (figure~\ref{fig:MetalLayers}b) is similar to that described in~\cite{S.CharlesDoret2012}: three layers of patterned aluminum insulated from one another by silicon dioxide. The bottom aluminum layer (M1) is grounded and prevents RF electric fields from penetrating into the RF-lossy silicon substrate.  The middle layer (M2) is patterned with control and RF electrodes as well as on-chip capacitors which reduce RF pickup on the control electrodes. The top metal layer (M3) is grounded and defines the boundary of the control electrodes. This simplifies the modeling  of trapping potentials and also helps to protect trap structures in M2 from physical damage~\cite{S.CharlesDoret2012}.  Seventy-eight control electrodes are arranged outside the RF electrodes in 50 $\mu$m wide rails with a 54 $\mu$m pitch (figure~\ref{fig:MetalLayers}a). The control electrodes in the corners of the junction are slightly larger to accommodate electrical leads. All gaps between electrodes are 4 $\mu$m wide. The electrodes between the RF rails are grounded by vias (located outside of the active region of the trap) to the chip ground plane below (M1). A 50 $\mu$m by 50 $\mu$m loading slot is etched through one of the center electrodes so that neutral atoms, supplied from an oven beneath the trap during trap loading, can reach the trapping region without electrically shorting the trap electrodes (figure~\ref{fig:MetalLayers}a,c).   RF electrode dimensions in the linear sections are chosen to establish the pseudopotential minimum at a height of 60 $\mu$m above the surface.  The rails are 40 $\mu$m wide and are separated by 80 $\mu$m (inner edge to inner edge). 

\begin{figure}[p]
\center 
\includegraphics[scale=1]{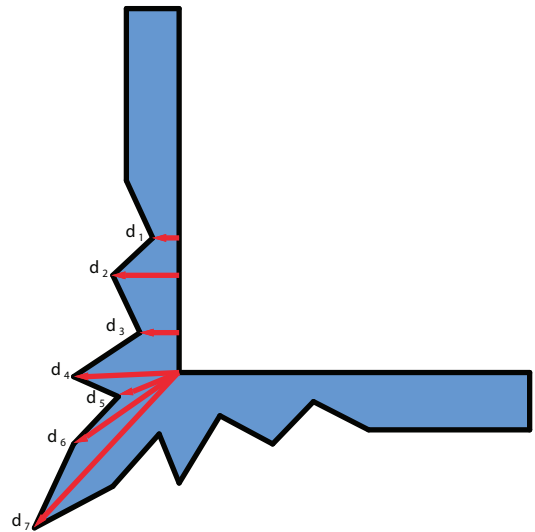} 
\caption{Control points for junction optimization. Each point is defined by a distance from the outer rail edge along a pre-defined direction, indicated here by the red arrows.}
\label{fig:RF_ElectrodeDesign}
\end{figure}

A junction naively assembled from the intersection of two linear sections does not provide adequate three-dimensional confinement to allow controlled transport~\cite{Wesenberg2008}. Therefore, we alter the RF electrode shape near the junction to increase trapping strength.  Working from an initial trial geometry we then optimize the shape of the rails to reduce the predicted ion heating rate during transport.  This is accomplished by placing seven control points along the inside edge of the RF electrode (figure~\ref{fig:RF_ElectrodeDesign}), giving seven degrees of freedom.  The locations of these points are modified with a genetic algorithm that employs an objective fitness function,
\begin{equation}
F =\int_0^{l_{max}} \left(\frac{\partial |\vec{E}_0\cdot \hat{l}|^2} {\partial l} \right) dl,\\
\label{eqs:FitnessFunction}	
\end{equation}
where the electric field due to application of the RF trap drive has the form
\begin{equation}
\vec{E}_{RF}\left(x,y,z,t\right)=\vec{E_0}\left(x,y,z\right)\cos(\Omega_{RF}t).
\label{eqs:ElectricField}
\end{equation}
\begin{figure}[p]
\center 
\includegraphics[scale=1]{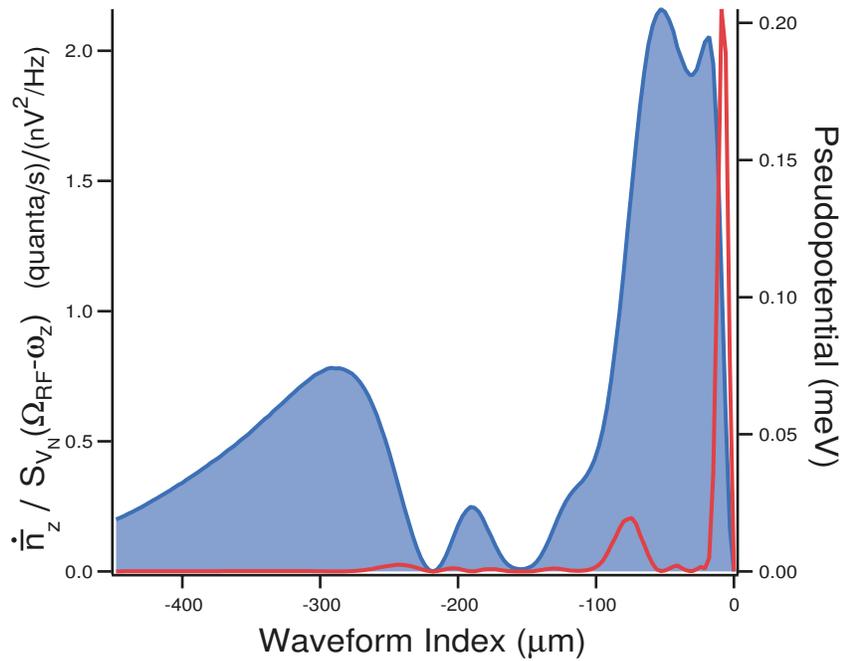} 
\caption{The ratio of the heating rate $\dot{\overline{n}}$ (in quanta/s) to voltage noise spectral density $S_{V_{N}}(\Omega_{RF}-\omega_{z})$, in units matching figure 8 of~\cite{Blakestad2011} (plotted in red). The shaded blue curve shows the trapping pseudopotential.  Both quantities are plotted versus the distance from the junction center along z (figure~\ref{fig:MetalLayers}a).}
\label{fig:HeatingRate}
\end{figure}
This fitness function is a measure of intrinsic secular heating along the z direction (figure~\ref{fig:HeatingRate}) during transport~\cite{Blakestad2009} due to spectral noise on the RF potentials.  The path $l$ follows the minimum of the pseudopotential as the ion is translated away from the center of the junction along one of the legs.  A candidate design is rejected if the pseudopotential is anticonfining in the direction perpendicular to the trap at any point along this path.  Each trial geometry generated by the genetic algorithm is evaluated by calculating the  field $\vec{E}_0$ with an in-house boundary element method (BEM) electrostatics solver, similar to those described in~\cite{Blakestad2011, KilianSinger2010}.

\section{Waveforms}

Ions are shuttled between different regions of the trap by applying transport waveforms~\cite{S.CharlesDoret2012}, which are smoothly varying sets of potentials applied to the control electrodes that produce a traveling harmonic well. In the linear regions of the trap, waveforms are designed for an axial secular frequency of $\omega_z = 2\pi\times 1$ MHz (for $^{40}$Ca$^+$). A 12.5$^\circ$ rotation of the radial secular axes from the trap normal ensures adequate Doppler cooling of all radial modes via lasers aligned parallel to the trap surface. Closer to the junction, the waveforms are designed to create a harmonic trapping potential with non-degenerate mode frequencies while minimizing sensitivity of the ion position to stray electric fields. 
\begin{figure}
\center 
\includegraphics[scale=1]{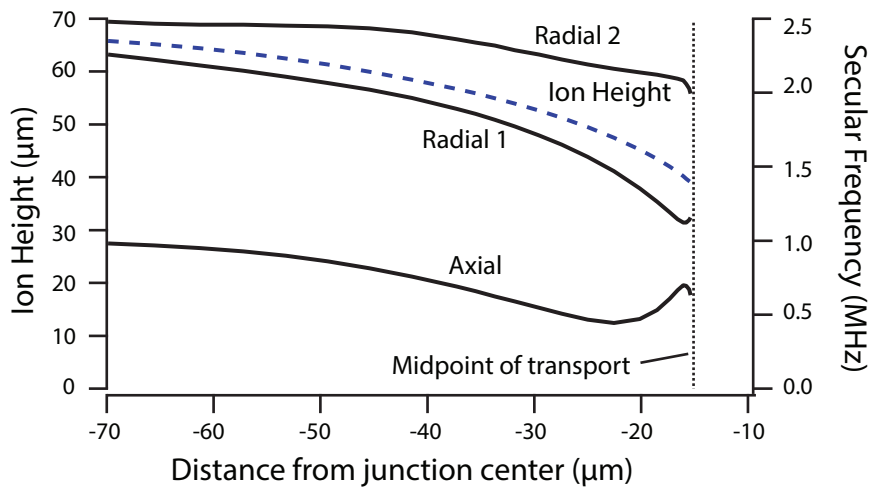} 
\caption{Calculated secular mode frequencies (solid) and ion height (dashed) for a waveform (-z$\rightarrow$+x or -x) designed for transport from one leg of the junction to the midpoint between the two legs. The potential forces the ion to circumnavigate the junction center at a 15 $\mu$m radius (see text).}
\label{fig:ModeHeight}
\end{figure} 

We can express the harmonic confinement in quadratic form,
\begin{equation}
	\Phi = \frac12 
		\left(\begin{array}{ccc} 
			x & y & z
		\end{array}\right)
		(M+Q)
		\left(\begin{array}{c}
			x\\ y\\ z		
		\end{array}\right),
	\label{eqs:HarmonicConfinement}
\end{equation}
where $\Phi$ is the trapping potential and M and Q are $3\times3$ matrices describing the control potentials and RF pseudopotential, respectively. Net confinement of the ion requires Tr(M+Q)$>$0. Poisson's equation enforces Tr(M)=0, hence the ion may only be trapped where Tr(Q)$>$0; control potentials can only redistribute the confinement provided by the RF pseudopotential and cannot increase confinement simultaneously in all three directions. Unfortunately, the RF pseudopotential confinement weakens significantly near the junction. Trapping the ion in this region thus requires using the control electrodes to share the weak pseudopotential confinement among all three directions (figure~\ref{fig:ModeHeight}), leading to small well-depths. We partially compensate for this by deliberately pushing the ion approximately 10 $\mu$m closer to the trap surface (figure~\ref{fig:ModeHeight}), increasing Tr(Q) at the expense of moving the ion away from the pseudopotential null and causing excess micromotion.  As such, the path actually followed by the ion is not the same as that followed in the junction optimization, likely causing the ion to experience more heating from RF noise than originally predicted.

To construct transport waveforms, we begin by expanding the harmonic portion of the pseudopotential in spherical harmonics for a series of locations along the desired path, calculating the eigen-axes of the pseudopotential alone.  Near the junction center the pseudopotential axes rotate sharply, and the associated eigen-frequencies become non-degenerate. Empirically we find that large control potentials are needed to rotate the secular axes away from the pseudopotential axes.  We therefore constrain the secular axes to closely overlap with the eigen-axes defined by the RF pseudopotential, allowing small deviations since doing so can increase the trap depth while staying within our control potential limits.  For each ion location we specify the following criteria: the height of the ion above the trap surface, any deviation in secular axes from the pseudopotential axes, one or more of the secular frequencies, and bounds of $\pm$8 V for the control potentials.  We use a simplex search over the space of control potentials to minimize a weighted, least-squares error function based on these criteria.  After calculating potential sets in this way for several locations along the desired path, we determine potentials for intermediate points by interpolating at 2 $\mu$m intervals.  Finally, we smooth the results to remove high spatial frequencies, as we have found that these do not improve the waveform but may contribute to heating during transport due to rapid swings in the control potentials. 

\begin{figure}
\center 
\includegraphics[scale=1]{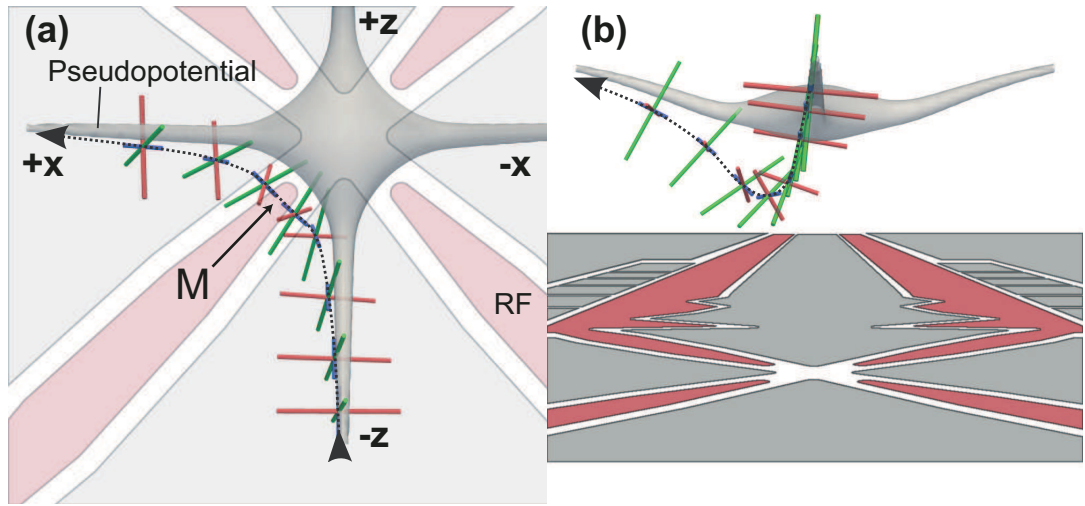} 
\caption{Ion path, pseudopotential isocontour, and secular mode axes for transport between two legs of the junction. (a) Top view, indicating the midpoint (M) of the transport waveform index as the ion circumnavigates the center and (b) a side view.}
\label{fig:ModePath}
\end{figure}  
 
Experimental characterization of the junction reveals a potential barrier at the junction center which is not predicted by our electrostatic models. This bump makes the confinement in the center of the junction sufficiently weak that stable transport directly through the center is difficult. We believe the barrier originates from incomplete etching of the oxide layer from the gaps at the center of the junction (where the gaps lie directly beneath the ion), leaving residual dielectric which can be charged by the Doppler cooling laser~\cite{S.CharlesDoret2012, MHarlander2010, ShannonX.Wang2011}.  To make transport more robust, we deliberately avoid a 15 $\mu$m radius around the junction center.  We also simulate the behavior of charged dielectric in the gaps by applying a positive potential to the M1 ground plane (figure~\ref{fig:MetalLayers}) in our electrostatic model.  This model reproduces the two effects we observe near the junction center: repulsion from the center, and barriers across the four RF spurs that project into the junction.  These barriers can cause double-wells to form during transport, leading to ballistic motion and associated heating as the ion moves between adjacent junction legs.  We adjust the potential applied to M1 in the model ($\sim$ 0.4 V) to roughly match the observed repulsion of the ion from the junction center and also calculate an adjustable correction potential that we may empirically tune to minimize the observed heating of the ion during transport.  The resulting calculated ion trajectory is shown in figure~\ref{fig:ModePath}.

We estimate that the transport waveform should be robust to stray fields of approximately 50 V/m without compromising the ion confinement. However, as previously characterized traps of similar construction exhibit stray fields of 100 V/m or more~\cite{S.CharlesDoret2012}, we generate additional compensation waveforms to null stray fields in each of the Cartesian directions at every point along the ion's trajectory. These compensation potentials are added to the transport waveform as needed empirically to minimize ion heating during transport.

\section{Characterization}

We characterize ion lifetime and transport by trapping $^{40}$Ca$^+$ in an apparatus similar to that described in~\cite{S.CharlesDoret2012}. National Instruments PXI-6733 16-bit DAC cards apply the transport waveforms. These cards apply voltage updates at 500 kHz; to reduce noise and associated ion heating, the control potentials are filtered by third-order Butterworth filters (60 kHz cut-off frequency) located just outside the vacuum chamber~\cite{Blakestad2011}. To cool all three modes of the ion in multiple legs of the junction, a Doppler cooling laser propagates at 45$^\circ$ to both the x and z directions. Fluctuations in the power of the fluorescence laser are stabilized to $<$1$\%$.
	
To characterize the junction we measure three figures of merit. First, we measure the lifetime of a stationary trapped ion without Doppler cooling, setting a lower bound on the ion loss-rate. We then explore the reliability of our transport waveforms by repeatedly transporting an ion through the junction, both with and without Doppler cooling.  This determines the rate of ion loss due to shuttling operations. Finally, we qualitatively explore the motional heating caused during transport by monitoring ion fluorescence as a function of position for truncated round-trip transports through the junction. 

\subsection{Ion lifetime}
\label{Subsection:IonLifetime}

\begin{figure}

\center 
\includegraphics[scale=1]{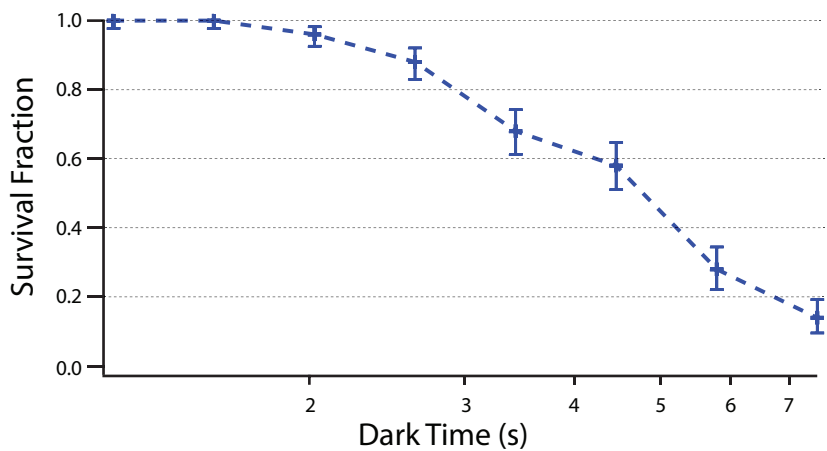} 
\caption{Ion survival fraction as a function of time without Doppler cooling. Each point is the accumulation of 50 experiments.}
\label{fig:Lifetime}
\end{figure}

The lifetime of a single ion trapped in one of the legs of the junction\footnote{$\Omega_{RF} = 58.55$~MHz, $V_{RF} = 91$~V$_{RMS}$, calcuated trap depth = 29~meV (axially limited)} is several hours when continuously Doppler cooled. Without cooling, the single-ion lifetime is approximately five seconds (figure~\ref{fig:Lifetime}), with a strongly non-exponential time dependence similar to that observed elsewhere~\cite{S.CharlesDoret2012}. This measurement is performed by repeatedly blocking the cooling laser for fixed periods of time and observing whether the ion remains trapped.  We performed the experiment fifty times for each fixed delay.

\subsection{Junction transport fidelity}
\label{Subsection:JunctionTransports fidelity}

To characterize the fidelity of shuttling through the junction, we perform 10$^6$ round-trip transports between two legs of the junction (-z$\rightarrow$+x$\rightarrow$-z) (figure~\ref{fig:MetalLayers}), traveling from a point 100~$\mu$m from the junction's center to a point 100~$\mu$m up the adjacent leg, and back.  The round-trip is executed in 200 total steps, requiring a time of 400~$\mu$s ($v_{ion}= 1$~m/s).  For the first 5$\times$10$^5$ transports we verify that the ion departs its initial position on the -z leg and then returns by monitoring ion fluorescence at the initial position using a PMT. We then shift the detection location to monitor the mid-point of the round-trip (in the +x leg) and verify the arrival and subsequent departure of the ion at this location.  Due to scatter of the fluorescence laser off of the complex topography near the junction, there is a non-negligible overlap between the fluorescence count histograms measured with and without an ion.  This limits our detection fidelity, and we can only place lower bounds on the transport reliability.  We confirm that the ion is not at an unexpected detection location with a probability of at least $99.8\%$, and that the ion arrives in the expected detection location at least $93\%$ of the time.

\subsection{Ion heating during junction transport}
\label{Subsection:JunctionHeating}

\begin{figure}[hbtp]
\center 
\includegraphics[scale=1]{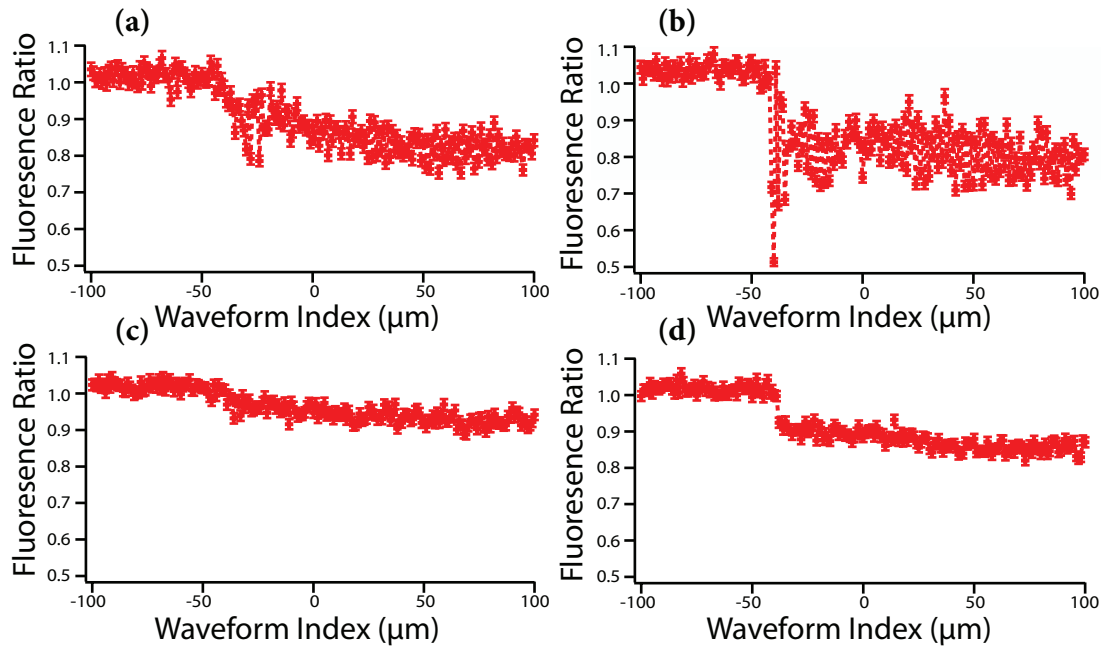} 
\caption{Ion heating due to transport as indicated by reduced ion fluorescence relative to a fully Doppler-cooled ion. Transport from (a) -z to the +x leg, (b) -z to -x, (c) +z to -x, and (d) +z to +x.  The localized fluorescence drop in (a) likely corresponds to secular mode frequencies moving into resonance with an unknown source of electric-field noise, while the sudden drops in (b) and (d) are likely due to the ion spilling between double wells formed during transport.  Each data point is an average of 1000 experiments.}
\label{fig:TransportHeating}
\end{figure}

Ion heating during transport can occur for two primary reasons. First, any discontinuous motion, or ``spilling" of the ion between adjacent potential wells, will heat the ion.  Such spilling behavior can occur should stray electric fields be present, as they may create multiple closely-spaced local minima in the weakly-confining transport waveform.  Second, the ion's motion can be driven by electrical noise that has frequency components near the trap secular frequencies. To distinguish between these two possibilities we execute a sequence of round-trip transports around the junction to one of the legs -- for example, along the path -z$\rightarrow$M$\rightarrow$+x (figure~\ref{fig:ModePath}) -- and compare the ion fluorescence before and after transport. To determine the spatial profile of any heating along a given path, the round-trip transport is truncated, with the ion pausing at an intermediate point for 10~$\mu$s before returning directly to the starting location.  Ion heating manifests as a reduction in the ion's fluorescence rate due to increased Doppler broadening. By comparing the ratio of fluorescence before and after transport, we produce a map of heating versus turning point location (figure~\ref{fig:TransportHeating}). Any discontinuity in the ion's motion due to spilling between potential wells should lead to a spatial discontinuity in the observed heating.  Paths along -z$\rightarrow$M$\rightarrow$+x and +z$\rightarrow$M$\rightarrow$-x (figure~\ref{fig:TransportHeating}a,c) show smooth reductions in ion fluorescence versus truncation point.  We infer that heating along these paths is due to noisy trapping potentials exciting the ion's secular motion rather than a discontinuity; the localized drop in figure~\ref{fig:TransportHeating}a likely corresponds to secular mode frequencies moving into resonance with an unknown source of electric-field noise.  In contrast, along the paths -z$\rightarrow$M$\rightarrow$-x and +z$\rightarrow$M$\rightarrow$+x (figure~\ref{fig:TransportHeating}b,d) there is a sharp step in fluorescence, suggesting a discontinuity in the transport waveform where the ion heats suddenly.  We believe this discontinuity is due to gradients in the stray fields present, preventing complete stray-field nulling in all junction legs using the common set of stray electric field compensations we applied for all four measurements.  However, additional tailoring of the transport waveform to fine-tune shuttling between these legs would likely eliminate this heating.  

\begin{figure}[hbtp]

\center 
\includegraphics[scale=1]{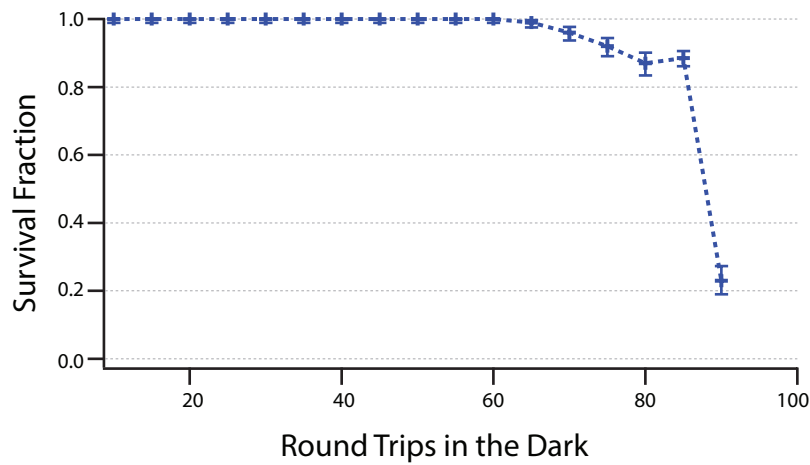} 
\caption{Ion survival fraction versus number of round-trip transports without cooling. Each data point represents 100 experiments. The fraction of experiments for which the ion returns to the original starting location gives the survival fraction.}
\label{fig:DarkTransports}
\end{figure}

Another measure of the heating induced by transport is given by the number of times that we can shuttle the ion back and forth through the junction without intermediate Doppler cooling between consecutive transports. By repeating this transport many times we determine the ion survival fraction as a function of the number of round-trip transports (figure~\ref{fig:DarkTransports}). We find that we can consecutively transport the ion through the junction sixty-five times with $>98\%$ reliability.  However, the survival fraction decreases sharply after approximately eighty-five round-trips, suggesting that ion loss is dominated by cumulative heating effects that increase the ion's energy beyond the trap depth.  This conjecture is supported by the fact that the eighty-five transports take approximately 34 ms, an interval far shorter than the ion lifetime without transport (approx. 5 s, figure~\ref{fig:Lifetime}).  We believe this heating is caused by noise on the RF and control potentials and by aliased  harmonics of the transport waveform~[13] around the 500~kHz DAC update rate.  The effects of these noise sources are exacerbated by the particularly low secular frequencies during the ion transport through the junction.  It should be possible to reduce or eliminate such heating by switching to DACs with update rates well above the maximum secular frequency and by improved filtering of the trapping potentials.  

\section{Conclusions}

We have designed, fabricated, and characterized a microfabricated X-junction surface-electrode ion trap, demonstrating reliable transport between the junction legs. More than 10$^5$ round-trip transports can be completed with intermediate cooling. Ion heating while shuttling through the junction is low enough to permit at least sixty-five consecutive round-trip transports without laser cooling, limited by electrical noise on the trapping potentials.  These results imply that an X-junction of similar design could be used to re-order a chain of ions or to shift ions between registers of a future large-scale trapped-ion quantum information processor.    

\section{Acknowledgments}
We would like to thank Kenton R. Brown for his comments on the manuscript.  This material is based upon work supported by the Georgia Tech Research Institute and the Office of the Director of National Intelligence (ODNI), Intelligence Advanced Research Projects Activity (IARPA) under U.S. Army Research Office (ARO) contract W911NF081-0315. All statements of fact, opinion, or conclusions contained herein are those of the authors and should not be construed as representing the official views or policies of IARPA, the ODNI, or the U.S. Government.


\addcontentsline{toc}{section}{References} 

\bibliography{reference_database}

\end{document}